\documentclass[10pt,final,letterpaper,twocolumn,twoside]{IEEEtran}
\usepackage{Steven}
\usepackage{amsmath}
\usepackage{amssymb}
\usepackage{graphicx} 

\newcommand{\vol}{\mathrm{vol}}
\newcommand{\Lbc}{\Lambda_{\rm bc}}
\newcommand{\Lsrs}{\Lambda_{\rm srs}}
\newtheorem{proofsketch}{Proof sketch}
\newtheorem{conjecture}{Conjecture}

\title{Geometric Approximations of Some Aloha-like Stability Regions}
\author{
\authorblockN{Nan Xie and Steven Weber\\}
\authorblockA{
Drexel University, Dept. of ECE, Philadelphia, PA 19104, USA\\
{\sf nx23@drexel.edu, sweber@ece.drexel.edu}
}}

\begin{document}

\maketitle

\begin{abstract}
Most bounds on the stability region of Aloha give necessary and sufficient conditions for the stability of an arrival rate vector under a specific contention probability (control) vector.  But such results do not yield easy-to-check bounds on the overall Aloha stability region because they potentially require checking membership in an uncountably infinite number of sets parameterized by each possible control vector.  In this paper we consider an important specific inner bound on Aloha that has this property of difficulty to check membership in the set.  We provide ellipsoids (for which membership is easy-to-check) that we conjecture are  inner and outer bounds on this set.  We also study the set of controls that stabilize a fixed arrival rate vector; this set is shown to be a convex set. 
\end{abstract}

\section{Introduction} 
\label{sec:1} 

This paper addresses geometric approximations of certain regions related to the stability of the slotted Aloha medium access control (MAC) protocol for a finite number of users.  The stability region of a protocol is the set of arrival rate vectors (with elements corresponding to exogenous arrival rates at each user's queue) such that each queue's length remains uniformly bounded.  This stability region is unknown for $n > 3$ users.  The $n=2$ user case was solved by Tsybakov and Mikhailov \cite{TsyMik1979} in 1979, and the $n=3$ user case was solved by Szpankowski \cite{Szp1994} in 1994 (the expressions for $n=3$ are rather unwieldy).  There is a large body of work giving inner and outer bounds on the stability region for general $n$ (some highlights include \cite{RaoEph1988,Ana1991,LuoEph1999,KomMaz2009sub}) but many of them are difficult to evaluate, as we now discuss.  

Let $\pbf \in [0,1]^n$ be the contention probability vector, where $p_i$ is the probability that user $i$ will transmit a packet in any time slot during which user $i$'s queue is non-empty at the start of that time slot, independently of anything else.  Let $\Lambda_A(\pbf)$ be the stability region of the Aloha protocol associated with a particular choice of $\pbf$, where $\xbf = (x_1,\ldots,x_n) \in \Lambda_A(\pbf)$ means the control $\pbf$ stabilizes each queue under arrival rates $\xbf$.  The Aloha stability region is the union of $\Lambda_A(\pbf)$ over all feasible control vectors $\pbf$:
\begin{equation}
\label{eq:1}
\Lambda_A = \bigcup_{\pbf \in [0,1]^n} \Lambda_A (\pbf).
\end{equation}
i.e., $\Lambda_A$ is defined by the parameterized regions $\Lambda_A(\pbf)$:
\begin{equation}
\Lambda_A = \left\{ \xbf : \exists \pbf \in [0,1]^n : \xbf \in \Lambda_A(\pbf) \right\}.
\end{equation}
One limitation of many of the Aloha stability bounds in the literature is that they do not yield easy-to-check necessary and sufficient conditions for stability, meaning (to the best of our knowledge) the literature {\em does not contain} results of the desired form:
\begin{eqnarray}
f(\xbf) \leq 0 & \Rightarrow & \xbf \in \Lambda_A \nonumber \\
g(\xbf) \leq 0 & \Rightarrow & \xbf \not\in \Lambda_A,
\end{eqnarray}
for given easily computable functions $f(\xbf),g(\xbf)$.  Instead, the literature contains results of the form:
\begin{eqnarray}
\tilde{f}(\xbf,\pbf) \leq 0 & \Rightarrow & \xbf \in \Lambda_A(\pbf) \nonumber \\
\tilde{g}(\xbf,\pbf) \leq 0 & \Rightarrow & \xbf \not\in \Lambda_A(\pbf),
\end{eqnarray}
for given easily computable functions $\tilde{f}(\xbf,\pbf),\tilde{g}(\xbf,\pbf)$ that each take as argument {\em both} $\xbf$ and $\pbf$.  These results, although very useful for testing stability membership in $\Lambda_A(\pbf)$ {\em for a specific control $\pbf$}, are of less use in testing stability membership overall in $\Lambda_A$ in \eqref{eq:1}.  Namely such results yield bounds of the form $\Lambda_{\tilde{f}} \subseteq \Lambda_A \subseteq \Lambda_{\tilde{g}}$ where
\begin{eqnarray}
\Lambda_{\tilde{f}} &=& \left\{ \xbf : \exists \pbf \in [0,1]^n : \tilde{f}(\xbf,\pbf) \leq 0 \right\} \nonumber \\
\Lambda_{\tilde{g}}^c &=& \left\{ \xbf : \forall \pbf \in [0,1]^n : \tilde{g}(\xbf,\pbf) \leq 0 \right\}.
\end{eqnarray}
The main motivation behind this paper is this: {\em most stability results on Aloha found in the literature are of limited value for evaluating stability of a rate vector $\xbf$ because it is not easy to test whether or not $\xbf \in \Lambda_{\tilde{f}}$ or $\xbf \in \Lambda_{\tilde{g}}$ since the parameter space $\pbf \in [0,1]^n$ defining these sets is uncountably infinite.}  Consequently, the goal of this paper is to develop inner and outer bounds on sets like $\Lambda_{\tilde{f}},\Lambda_{\tilde{g}}$ that are easy to evaluate.

The paradigmatic example we study in this paper is the set 
\begin{equation}
\boxed{
\Lambda \equiv \left\{ \xbf : \exists \pbf \in [0,1]^n : x_i \leq p_i \prod_{j \neq i} (1-p_j), ~ i \in [n] \right\}.}
\end{equation}
The expression $p_i \prod_{j \neq i} (1-p_j)$ is the worst-case service rate for user $i$'s queue, meaning the  service rate when all users have non-empty queues and are therefore eligible for channel contention.  In particular, user $i$'s packet is successful in such a time slot if user $i$ elects to contend (with probability $p_i$) and each other user $j \neq i$ does not contend (each with independent probability $1-p_j$ for non-empty queue).  In fact $\Lambda \subseteq \Lambda_A$, since an arrival rate that is stabilized under a worst-case service rate is certainly stabilized under a better service rate.  Further, for $n=2$ we have $\Lambda = \Lambda_A$ \cite{TsyMik1979}.  The set $\Lambda$ has been addressed in the literature in \cite{MasMat1985,Pos1985,Ana1991}.   Massey and Mathys \cite{MasMat1985} showed $\Lambda$ is the capacity region of the collision channel without feedback, Post \cite{Pos1985} showed that the complement of $\Lambda$ in $\Rbb_+^n$ is convex (thus $\Lambda$ itself is non-convex) and characterized its tangent hyperplane at each point on its boundary, and Anantharam \cite{Ana1991} showed $\Lambda$ {\em is} the stability region for Aloha for a certain correlated arrival process. Furthermore, $\Lambda$ is widely conjectured to coincide with \textit{the} Aloha stability region $\Lambda_{A}$ for general arrival processes \cite{TsyMik1979, RaoEph1988, LuoEph2006}. Recently, by using \textit{mean field analysis} and assuming each queue's evolution is independent, Bordenave et al. \cite{BorMcD2008} were able to show that \textit{asymptotically} this conjecture is true. 

Our approach is to provide simple geometric constructions that both inner bound and outer bound $\Lambda$, where it is straightforward to check membership in both bounds (note it is not easy to test membership in $\Lambda$).  Our figure of merit in evaluating these constructions is the volume of the set relative to the volume of $\Lambda$.  In order to do so we first give in \S\ref{sec:2} a new result characterizing $\vol(\Lambda)$.  Next, in \S\ref{sec:3} we give a simple inner bound and a simple (conjectured) outer bound on $\Lambda$ for arbitrary $n$ that are inspired by the geometry of $\Lambda$ for $n=2$, but these bounds are loose in that their volumes poorly approximate the volume of $\Lambda$ as $n$ increases. Then \S\ref{sec:4} contains our primary result on geometric approximations of $\Lambda$.  Namely, we give (conjectured) ellipsoid constructions $\Emc_I,\Emc_O$ such that $\Lambda_I \subseteq \Lambda \subseteq \Lambda_O$ where $\Lambda_I = \Smc \setminus \Emc_I$ and $\Lambda_O = \Smc \setminus \Emc_O$, and $\Smc = \{ \xbf \geq \mathbf{0} : \langle \xbf, \mathbf{1} \rangle \equiv \sum_{i=1}^{n} x_{i} \leq 1 \}$ is the simplex.  

In \S\ref{sec:5} we change our focus to a different set relevant to Aloha.  Namely, given an arrival rate vector $\xbf$, define the set of stabilizing controls $\Pmc(\xbf)$ assuming a worst-case service rate:
\begin{equation}
\label{eq:t}
\Pmc(\xbf) = \left\{ \pbf \in [0,1]^n : x_i \leq p_i \prod_{j \neq i} (1-p_j), ~ i \in [n] \right\}.
\end{equation}
This set is related to $\Lambda$ in that $\Pmc(\xbf) = \emptyset$ iff $\xbf \not \in \Lambda$.  Whereas $\Lambda$ is an important bound for the stability region $\Lambda_A$, the set $\Pmc(\xbf)$ is important from a more practical operational perspective: given a desired arrival rate vector $\xbf$, it is an inner bound on the control options available to the network administrator.  We establish that $\Pmc(\xbf)$ is a convex set.  We offer a brief conclusion in \S\ref{sec:6}.  

\section{Volume of the set $\Lambda$} 
\label{sec:2} 

In this section we give an expression for the volume of $\Lambda$.  


\begin{proposition}
\label{pro:1}
The set $\Lambda$ defined in \eqref{eq:1} has volume $\vol(\Lambda) = $
\begin{equation}
\label{eq:2}
\sum_{\kbf \in \Kmc} \binom{n-2}{\kbf} \prod_{S \subseteq [n]} (-1)^{|S| k_S} 
\frac{\prod_{i=1}^n \left( \sum_{S \ni i} k_S \right)!}{\left(n + 1 + \sum_{S \subseteq [n]} |S| k_S \right)!}
\end{equation}
where $\binom{n-2}{\kbf} \equiv \binom{n-2}{k_S, S \subseteq [n]}$ is a multinomial coefficient, and 
\begin{equation}
\Kmc = \left\{ \kbf = (k_S, S \subseteq [n]) \geq \mathbf{0}~ : ~ |\kbf| \equiv \sum_{S \subseteq [n]} k_S = n-2 \right\}.
\end{equation} 
\end{proposition}


\begin{proofsketch}
Define the simplex $\Smc \equiv \{ \pbf \geq \mathbf{0} : \langle \pbf, \mathbf{1} \rangle \leq 1 \}$ and $\rho(\pbf) \equiv \prod_{i=1}^n (1-p_i)$. 
One can show that the mapping $\pbf \mapsto \xbf$ given by
\begin{equation}
\label{eq:o}
x_i(\pbf) = p_i \prod_{j \neq i}(1-p_j), ~ i \in [n],
\end{equation}
is a bijection from $\Smc$ to $\Lambda$.  In fact this function is also a bijection from $\partial \Smc \equiv \left\{ \pbf \geq \mathbf{0}: \langle \pbf, \mathbf{1} \rangle = 1 \right\}$ (which forms a ``face'' on the boundary of $\Smc$) to 
\begin{equation}
\label{eq:g}
\partial \Lambda \equiv \left\{ \xbf : \exists \pbf \in \partial \Smc : x_i = p_i \prod_{j \neq i}(1-p_j), ~ i \in [n] \right\}
\end{equation}
(which forms a ``face'' on the boundary of $\Lambda$).    
Let $\tilde{\Jbf}(\pbf) \equiv \Jbf(\pbf)/\rho(\pbf)$, where $\Jbf(\pbf)$ is the Jacobian for this mapping. 
The fact that $\det(\alpha A) = \alpha^n \det A$ for any scalar $\alpha$ and any $n \times n$ matrix $A$ yields $\det \Jbf(\pbf) = \rho(\pbf)^n \det \tilde{\Jbf}(\pbf)$.  Abramson \cite{Abr1977} showed that $\rho(\pbf)^2 \det \tilde{\Jbf}(\pbf) = 1 - \langle \pbf, \mathbf{1} \rangle$, which gives $\det \Jbf(\pbf) = \rho(\pbf)^{n-2} \left(1 - \langle \pbf, \mathbf{1} \rangle \right)$.  Substituting this into the general expression for volume  $\mathrm{vol}(\Lambda) = \int_{\Smc} \det \Jbf(\pbf) \drm \pbf$ shows $\mathrm{vol}(\Lambda) = \int_{\Smc} \prod_i (1-p_i)^{n-2} (1-\sum_i p_i) \drm \pbf$.

To get a better closed-form expression, we leverage results in  \cite{GruMol1978} (2.3) on integration of certain functions over a simplex: 
\begin{equation}
\label{eq:a}
\int_{\Smc} \pbf^{\boldsymbol\alpha} \left(1 - \sum_i p_i \right)^{\alpha_0} \drm \pbf = \frac{\prod_{i=0}^n \alpha_i !}{\left(n+\sum_{i=0}^n \alpha_i \right)!},
\end{equation}
where $\Smc$ is the standard simplex, $\pbf = (p_1,\ldots,p_n)$, $\boldsymbol\alpha = (\alpha_1,\ldots,\alpha_n)$, and $\pbf^{\boldsymbol\alpha} = \prod_{i=1}^n p_i^{\alpha_i}$. Defining $p_S = \prod_{i \in S} p_i$, the multinomial theorem gives:
\begin{equation}
\left( \prod_i (1-p_i) \right)^{n-2} = \sum_{\kbf \in \Kmc} \binom{n-2}{\kbf} \prod_{S \subseteq [n]} \left( (-1)^{|S|} p_S \right)^{k_S}.
\end{equation}
Next:
\begin{equation}
\prod_{S \subseteq [n]} p_S^{k_S} = \prod_{i=1}^n p_i^{\alpha_i(\kbf)}, ~ \alpha_i(\kbf) = \sum_{S \ni i} k_S, ~ i \in [n],
\end{equation}
Substituting all these into $\mathrm{vol}(\Lambda)$ yields the proposition.
$\blacksquare$
\end{proofsketch}

Note that $|\Kmc|=1$ for $n=2$ and for any $n > 2$ we have
\begin{equation}
|\Kmc| = \sum_{j=1}^{n-2} \binom{2^n}{j} \binom{j+(n-2-j)-1}{n-2-j},
\end{equation}
where $j$ is the number of non-zero components of a given $\kbf \in \Kmc$. Unfortunately $|\Kmc|$ grows super-exponentially in $n$, posing huge computational overhead for even moderate $n$.

\section{Preliminary inner and outer bounds on $\Lambda$} 
\label{sec:3} 

\begin{figure}[!htbp]
\centering
\includegraphics[width=3.5in]{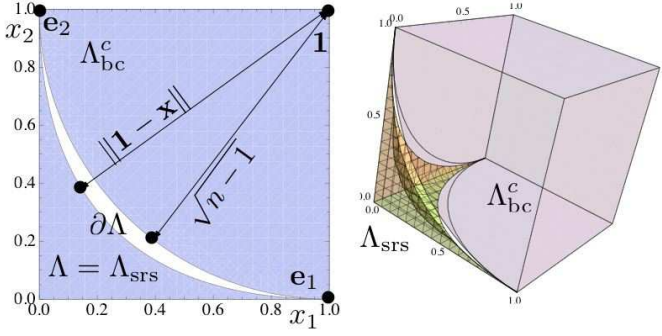}
\caption{The $\Lsrs$ inner bound and $\Lbc$ outer bound for $n=2$ (left) and $n=3$ (right). Note the labels indicate the {\em complement} of $\Lbc$, i.e., $b(\mathbf{1},\sqrt{n-1}) \cap [0,1]^n$.  The $\Lsrs$ bound is tight for $n=2$.}
\label{fig:1}
\end{figure}

The Aloha stability region for $n=2$ from \cite{TsyMik1979} is:
\begin{equation}
\Lambda_A = \Lsrs \equiv \left\{ \xbf \geq \mathbf{0}  : \sqrt{x_1} + \sqrt{x_2} \leq 1 \right\},
\end{equation}
where the notation $\Lsrs$ will be explained shortly.  It is straightforward to establish that $\Lambda_A = \Lambda$ for $n=2$.  In particular, for $n=2$, testing membership in $\Lambda = \Lambda_A$ is trivial since there is no parameterized set that must be potentially checked for each $\pbf \in [0,1]^2$.  Further, the shape of $\Lambda$ suggests it is well-approximated by the complement of a ball, namely $\Lbc \equiv \Smc \setminus b(\mathbf{1},1)$, where $b(\xbf_o,r) \equiv \{ \xbf : \|\xbf-\xbf_o\| \leq r\}$ is the ball centered at $\xbf_o$ of radius $r$.  Fig.~\ref{fig:1} (left) shows $\Lsrs = \Lambda = \Lambda_A \subseteq \Lbc$ for $n=2$.  

These results for $n=2$ motivate the following quantities, defined for arbitrary $n$:
\begin{eqnarray}
\label{eq:c}
\Lsrs &=& \left\{ \xbf \geq \mathbf{0} : \sum_{i=1}^n \sqrt{x_i} \leq 1 \right\} \nonumber \\
\Lbc &=& \left\{ \xbf \in [0,1]^n : \|\xbf - \mathbf{1}\| \geq \sqrt{n-1} \right\}.
\end{eqnarray}
The labels ``srs'' and ``bc'' stand for {\em square root sum} and {\em ball complement}, respectively.  The latter name is appropriate since $\Lbc = [0,1]^n \setminus b(\mathbf{1},\sqrt{n-1})$, see Fig.~\ref{fig:1}. 

\begin{proposition}
\label{pro:2}
The set $\Lsrs$ in \eqref{eq:c} is an inner bound for $\Lambda$, i.e., $\Lsrs \subseteq \Lambda$ for $n \geq 2$.
\end{proposition}

\begin{proofsketch}
Fix a point $\xbf' \in \Lsrs$.  Due to \textit{coordinate convexity} of $\Lambda,\Lsrs$, it suffices to produce a point $\xbf \in \Lambda$ with $x_i \geq x_i'$ for each $i \in [n]$.  Set $\pbf$ with $p_i = \sqrt{x_i'}$ for each $i$ and set $\xbf$ with $x_i = p_i \prod_{j \neq i} (1-p_j)$ for each $i$.  Clearly $\xbf \in \Lambda$.  It remains to show $x_i \geq x_i'$ for each $i$.  Note $\xbf' \in \Lsrs$ ensures $\sum_i p_i \leq 1$.  Define independent events $A_1,\ldots,A_n$ with $\Pbb(A_i) = p_i$ for each $i \in [n]$.  It follows that
\begin{equation}
1 - \Pbb \left( \bigcup_{j \neq i} A_j \right) = \Pbb \left( \bigcap_{j \neq i} A_j^c \right)  = \prod_{j \neq i} (1-p_j).
\end{equation}
Then for any $i$, by the union bound:
\begin{equation}
x_i = p_i \left( 1 - \Pbb \left( \bigcup_{j \neq i} A_j \right) \right) \geq  p_i \left( 1 - \sum_{j \neq i} \mathbb{P} (A_j)\right) \geq p_i^2, 
\end{equation} 
and thus $x_i \geq x_i'$. $\blacksquare$
\end{proofsketch}

\begin{conjecture}
\label{con:1}
The set $\Lbc$ in \eqref{eq:c} is an outer bound for $\Lambda$, i.e., $\Lambda \subseteq \Lbc$ for $n \geq 2$.
\end{conjecture}

At this point we only provide a possible methodology towards proving this conjecture. For any $\xbf \in \Lambda$ we must show that $\xbf \in \Lbc$, meaning $\|\xbf - \mathbf{1}\| \geq \sqrt{n-1}$.  To show this it suffices to show $\min_{\xbf \in \Lambda} \|\mathbf{1} - \xbf \| \geq \sqrt{n-1}$, which is equivalent to showing
\begin{equation}
\min_{\pbf \in [0,1]^n} f(\pbf) \equiv \sum_i \left(1 - p_i \prod_{j \neq i}(1-p_j) \right)^2 \geq n-1.
\end{equation}
The approach is based on how many distinct non-zero values the components of $\pbf$ assume. $i)$ If $\pbf$ has only one distinct non-zero value then it is easy to see $f(\ebf_i) = n-1$ for each of the $n$ unit vectors $\ebf_1,\ldots,\ebf_n$ (where $\ebf_i$ has a one in position $i$ and zero in all the other positions). Lemma \ref{lem:1} states this value is optimal over the class of all ``quasi-uniform'' vectors addressed by the lemma. $ii)$ If $\pbf$ has 2 or more distinct non-zero values in all its component positions, it is hoped to show that such a $\pbf$ can not be a \textit{global} minimizer either because $f(\pbf) > n-1$ or because $\pbf$ violates the Karush-Kuhn-Tucker (KKT) conditions required for optimality (In this case ``regularity'' is guaranteed, which justifies using KKT).  

%

\begin{lemma}
\label{lem:1}
Suppose $\pbf$ has $k$ non-zero elements, each equal to $1/k$ and $n-k$ zero elements.  Then $f(\pbf) \geq n-1$, with equality only when $k=1$.
\end{lemma}

\begin{proofsketch}
Note the aforementioned bijection property allows us to let $\sum_{i}p_{i}=1$. If $k=1$, $f(\pbf) \geq n-1$ is trivially seen to hold. If $k \geq 2$, showing $f(\pbf) \geq n-1$ is equivalent to showing $\left(1-k^{-1}\right)^{k-1} \leq  k \left(1-\sqrt{1-\frac{1}{k}} \right)$.  The derivative of its LHS can be shown to be negative using the inequality $\log y \leq y-1$. Thus the sequence $\{(1-1/k)^{k-1}\}$ is upper bounded by $ \left(1-\frac{1}{2}\right)^{2-1} = \frac{1}{2}$. On the other hand, using AM-GM inequality $\sqrt{1-1/k} \leq (1-1/k+1)/2$ (equality cannot hold when $k\geq 2$), one can see the above RHS is lower bounded by $\frac{1}{2}$.
$\blacksquare$
\end{proofsketch}

As some further remark about the $\Lbc$ bound conjecture, we note Cauchy-Schwarz gets close but is insufficient by itself:
\begin{equation*}
\sum_i (1-x_i)^2 \geq \frac{(\sum_i (1-x_i))^2}{n} = \frac{(n - \sum_i x_i)^2}{n} \geq \frac{(n-1)^2}{n}.
\end{equation*}
The next proposition gives the volume of $\Lsrs$. $\Lbc$ extends beyond the positive orthant so calculating its volume requires a difficult accounting of intersecting polar caps.

\begin{proposition}
\label{pro:3}
The volume of the inner bound $\Lambda_{srs}$ is
\begin{equation}
\label{eq:u}
\vol(\Lsrs) = \frac{2^n}{(2n)!} \leq \vol(\Lambda).
\end{equation}
\end{proposition}

\begin{proofsketch}
Use change of variable $y_i = \sqrt{x_i}$ for $i \in [n]$, and define $s_k = \sum_{i=k}^n y_i$ so that the volume becomes
\begin{equation}
\label{eq:d}
\vol(\Lsrs) = 2^n \int_{[0,1]^n}  \mathbf{1}_{s_1 \leq 1} y_1 \drm y_1 \cdots y_n \drm y_n. 
\end{equation}
Integration gives:
\begin{equation}
\label{eq:e}
\frac{\int_0^{1-s_{k+1}} ((1-s_{k+1})-y_k)^{2k-2} y_k \drm y_k}{(1-s_{k+1})^{2k}} = \sum_{i=0}^{2k-2} \frac{(-1)^i}{i+2} \binom{2k-2}{i} 
\end{equation}
for each $k$.  By induction on \eqref{eq:d} and \eqref{eq:e} we can establish:
\begin{equation}
\vol(\Lsrs) = 2^n \prod_{k=1}^n \left( \sum_{i=0}^{2k-2} \frac{(-1)^i}{i+2} \binom{2k-2}{i} \right).
\end{equation}
It is straightforward to show that
\begin{equation}
\sum_{i=0}^{2k-2} \frac{(-1)^i}{i+2} \binom{2k-2}{i} = \frac{1}{(2k)(2k-1)}, ~\prod_{k=1}^n (2k)(2k-1) = (2n)!
\end{equation}$\blacksquare$
\end{proofsketch}

As shown in Fig.~\ref{fig:vol}, $\vol(\Lsrs)$ scales rather poorly in $n$ relative to $\vol(\Lambda)$, therefore we seek improved bounds for $\Lambda$.

\section{Ellipsoid inner and outer bounds on $\Lambda$} 
\label{sec:4} 

\begin{figure}[!ht]
\centering
\includegraphics[width=3.0in]{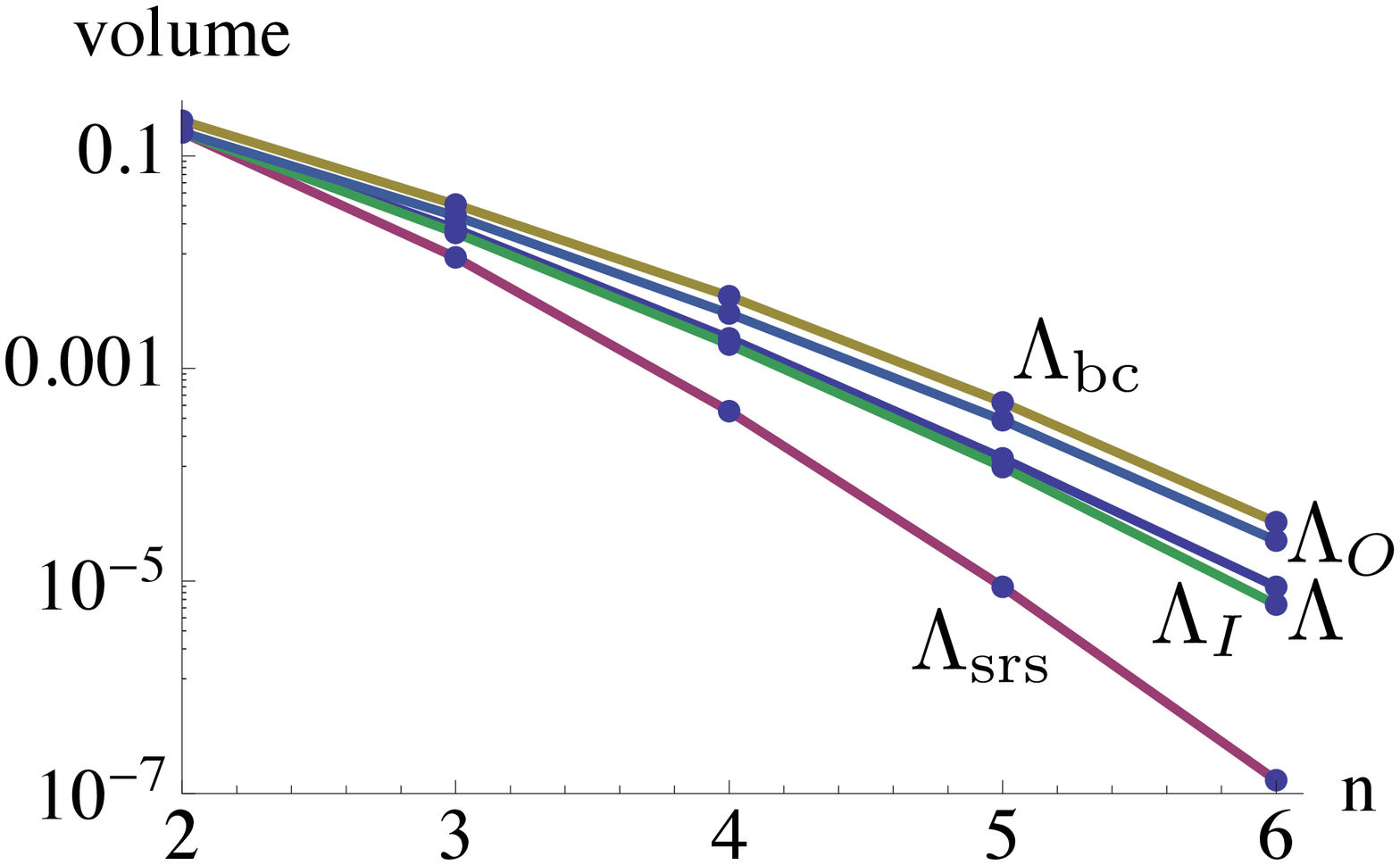}
\caption{The volumes of $\Lambda,\Lsrs,\Lbc,\Lambda_I,\Lambda_O$ versus $n$.}
\label{fig:vol}
\end{figure}

The fact that the $\Lbc$ curve is superior to the $\Lsrs$ curve in Fig.~\ref{fig:vol} suggests the complement of $\Lambda$ may be well approximated by appropriately chosen ellipsoids.  To this end, we consider ellipsoids $\Emc$ of the form (\cite{BoyVan2004}):
\begin{equation}
\label{eq:p}
\Emc = \left\{ \xbf : (\xbf - \cbf)^{\Tsf} \Rbf^{-1} (\xbf - \cbf) \leq 1 \right\}.
\end{equation}
Here $\cbf$ is the center of the ellipsoid and the $n \times n$ matrix $\Rbf$ has the spectral decomposition $\Rbf = \Qbf \Dbf \Qbf^{-1}$ where $\Qbf = [\qbf_1 \cdots \qbf_n]$ is orthonormal and holds the eigenvectors of $\Rbf$ which are the directions of the $n$ axes of the ellipsoid, and $\Dbf = \mathrm{diag}(d_1,\ldots,d_n)$ holds the eigenvalues of $\Rbf$, and each $a_i = \sqrt{d_i}$ is the semi-axis length  in the direction $\qbf_i$.


We assert that the symmetries in the set $\Lambda$ suggest symmetries for $\Emc$.  In particular, the center $\cbf$ should lie on the ray $(1,\ldots,1)$, i.e., $\cbf = c \mathbf{1} = (c,\ldots,c)$ for some scalar $c > 1/n$ to be determined.  Further, one axis of the ellipsoid should point along this same ray, i.e., have associated unit vector $\qbf_1 = \mathbf{1} = (1,\ldots,1)/\sqrt{n}$ and associated length $a_1$.  
Imagine we rotate the coordinate system so that the vector $\ebf_1$ is aligned with the vector $\mathbf{1}$.  Then this induces a permutation symmetry among the remaining rotated $n-1$ coordinates.  Hence, the ellipsoid is spherical in those directions (meaning $a_2 = \cdots = a_n$), leaving three degrees of freedom: $c,a_1,a_2$.

Our approach is to approximate the surface $\partial \Lambda$ with part of the surface of an ellipsoid placed in the complemenet of $\Lambda$, and then form inner and outer bounds on $\Lambda$ by subtracting these ellipsoids from the unit simplex $\Smc$.  The following conjecture defines the proposed ellipsoids and the corresponding regions that we believe to be inner and outer bounds on $\Lambda$.

Define the ellipsoids $\Emc_I,\Emc_O$ with center $\cbf = (c,\ldots,c)$ and $\Rbf = \Qbf \Dbf \Qbf^{-1}$ with orthonormal $\Qbf = (\qbf_1,\ldots,\qbf_n)$ for $\qbf_1 = (1,\ldots,1)/\sqrt{n}$ and $\qbf_2,\ldots,\qbf_n$ otherwise arbitrary.  The two distinct semi-axes lengths are:
\begin{equation} \label{eq:EIEO}
\begin{array}{lll}
\Emc_I: &  a_1^2 = n \left(c - \frac{1}{n}\left(1-\frac{1}{n}\right)^{n-1} \right)^2,  & a_2^2 = \frac{(n-1) a_1^2}{n a_1^2  - (nc-1)^2} \\ 
\Emc_O: & a_1^2 = (nc-1) c, & a_2^2 = (n-1)c
\end{array}.
\end{equation}
Define also the regions $\Lambda_I \equiv \Smc \setminus \Emc_I$ and $\Lambda_O \equiv \Smc \setminus \Emc_O$.  

\begin{conjecture}
\label{con:2}
The regions $\Lambda_I,\Lambda_O$ form inner and outer bounds on $\Lambda$: $\Lambda_I \subseteq \Lambda \subseteq \Lambda_O$ for $n \geq 2$.
\end{conjecture}

The expressions for $a_1,a_2$ for $\Emc_I,\Emc_O$ are derived essentially based on Lemma \ref{invarianceofR}, Corollary \ref{passingei}, and Lemma \ref{commontangency} given below. In words, $\Emc_{I}$ is such that it passes through $\ebf_1,\ldots,\ebf_n$ and the ``all-rates-equal'' point $\mbf = m \mathbf{1} \in \partial \Lambda$ where $m = \frac{1}{n}(1-\frac{1}{n})^{n-1}$.  It can be shown that $\partial \Emc_I$ is tangent at $\mbf$ with $\partial \Lambda$. $\Emc_{O}$ is such that it passes through each $\ebf_i$ and further $\partial \Emc_O$ is tangent with $\partial \Lambda$ at each $\ebf_i$. Note when $c=1$, $\Lambda_O = \Lambda_{bc}$ since in this case $\Emc_O$ is a ball with $a_1 = a_2 = \sqrt{n-1}$.

%


The following lemma says that, under given conditions, the ellipsoid is invariant to the choice of $\qbf_{2}, \ldots, \qbf_{n}$ provided $\Qbf$ is orthonormal and $\qbf_1 = \mathbf{1}/\sqrt{n}$. 

\begin{lemma} \label{invarianceofR}
For any ellipsoid in the form of \eqref{eq:p}, if $\qbf_1 = (1,\ldots,1)/\sqrt{n}$ and $\Dbf = \mathrm{diag}(a_1^2,a_2^2,\ldots,a_2^2)$, then $\Rbf^{-1} = \zeta \mathbf{1} + a_2^{-2} \mathbf{I}$, where $\zeta \equiv \frac{1}{n} \left(a_1^{-2}-a_2^{-2} \right)$, $\mathbf{1}$ is the $n \times n$ matrix with $\mathbf{1}_{ij}=1$ for each $i,j$ and $\Ibf$ is the identity matrix.
\end{lemma}

The proof is straightforward and omitted. This simple form of $\Rbf^{-1}$ allows us to characterize the ellipsoid that passes through each $\ebf_i$:

\begin{corollary} \label{passingei}
For any $c > 1/n$ and $a_1 > \sqrt{n}(c-1/n)$, setting 
\begin{equation}
a_2^2 = \frac{(n-1) a_1^2}{n a_1^2  - (nc-1)^2}
\end{equation}
ensures $\ebf_i \in \partial \Emc$ for $i \in [n]$.
\end{corollary}

The next lemma tells what is needed for $\partial \Emc$ and $\partial \Lambda$ to share a common tangent point:
\begin{lemma} 
\label{commontangency}
Define $\qbf_{t} \equiv \mathbf{1} - \pbf_{t}$, $\partial \Emc$ and $\partial \Lambda$ share a point of tangency at $\xbf_t$ if for each $i \in [n-1]$:
\begin{equation}
\left( \frac{a_{2}}{a_{1}}\right)^{2} = \frac{(q_{t,i}-q_{t,n}) \Sigma_{j=1}^{n} x_{t,j} + n (q_{t,n} x_{t,i} - q_{t,i} x_{t,n})}{(q_{t,i}-q_{t,n})(\Sigma_{j=1}^{n} x_{t,j} - nc)}.
\end{equation}
\end{lemma}

\begin{proofsketch}
Using the bijection \eqref{eq:o} between $\partial \Lambda$ and $\partial \Smc$ it follows that \cite{MasMat1985} there is a unique $\pbf_t \in \partial \Smc$ with $\langle \pbf, \mathbf{1} \rangle = 1$ corresponding to $\xbf_t$.  
Post \cite{Pos1985} established that the tangent hyperplane to $\Lambda$ at a point $\xbf_t \in \partial \Lambda$ with $\xbf_t \mapsto \pbf_t$ is $\langle \mathbf{1}-\pbf_t, \xbf \rangle = \rho(\pbf_t)$, for $\rho(\pbf_t) = \prod_i (1-p_{t,i})$.  Next, using the implicit function theorem it is straightforward to establish that the tangent hyperplane to an arbitrary ellipsoid $\Emc$ \eqref{eq:p} at a point $\xbf_t \in \partial \Emc$ is given by $\langle \wbf, \xbf_{t} \rangle = b$ where 
\begin{equation}
\label{eq:q}
\wbf = - \frac{\Rbf^{-1}(\xbf_t - \cbf)}{\Rbf^{-1}_n(\xbf_t - \cbf)}, ~ b = \langle \wbf, \xbf_t \rangle .
\end{equation}
Here $\Rbf^{-1}_n$ is the $n^{th}$ row of $\Rbf^{-1}$. 
Finally equating (after appropriate normalization) the components of the normal vector of $\partial \Lambda$ with those of $\partial \Emc$ yields this lemma.  $\blacksquare$ 
\end{proofsketch}

\begin{figure}[!ht]
\centering
\includegraphics[width=1.7in]{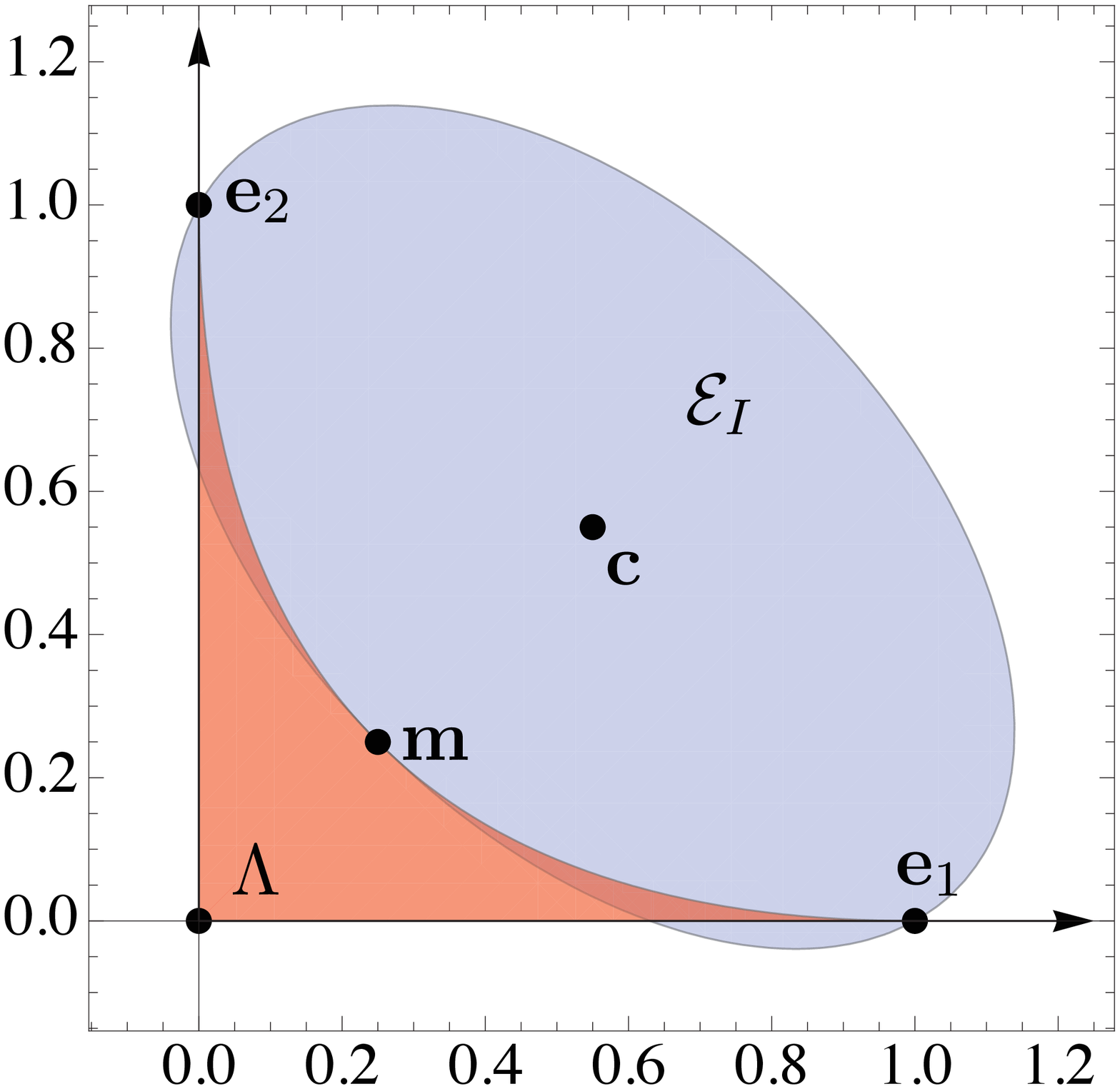}
\includegraphics[width=1.7in]{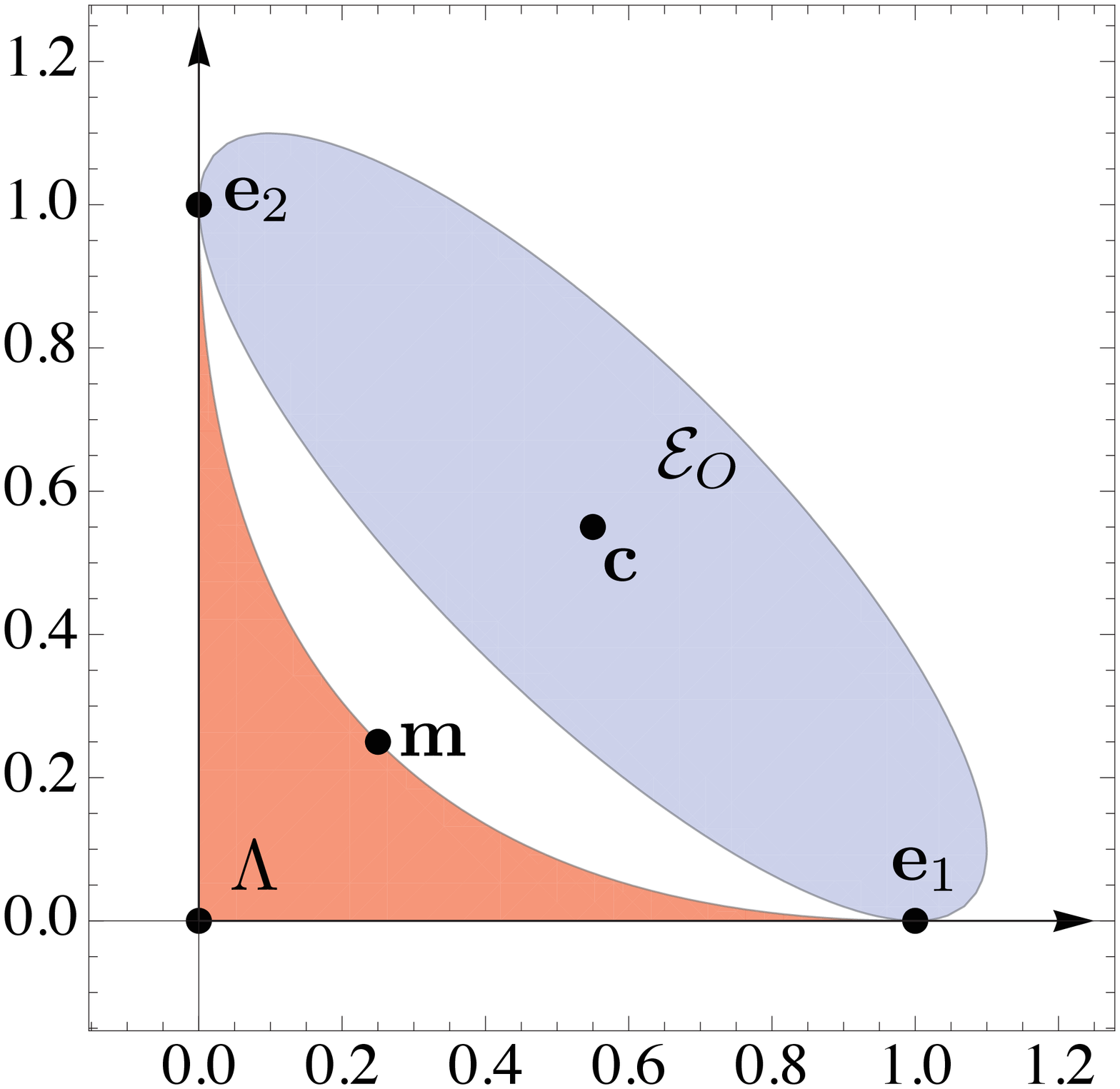}
\includegraphics[width=1.7in]{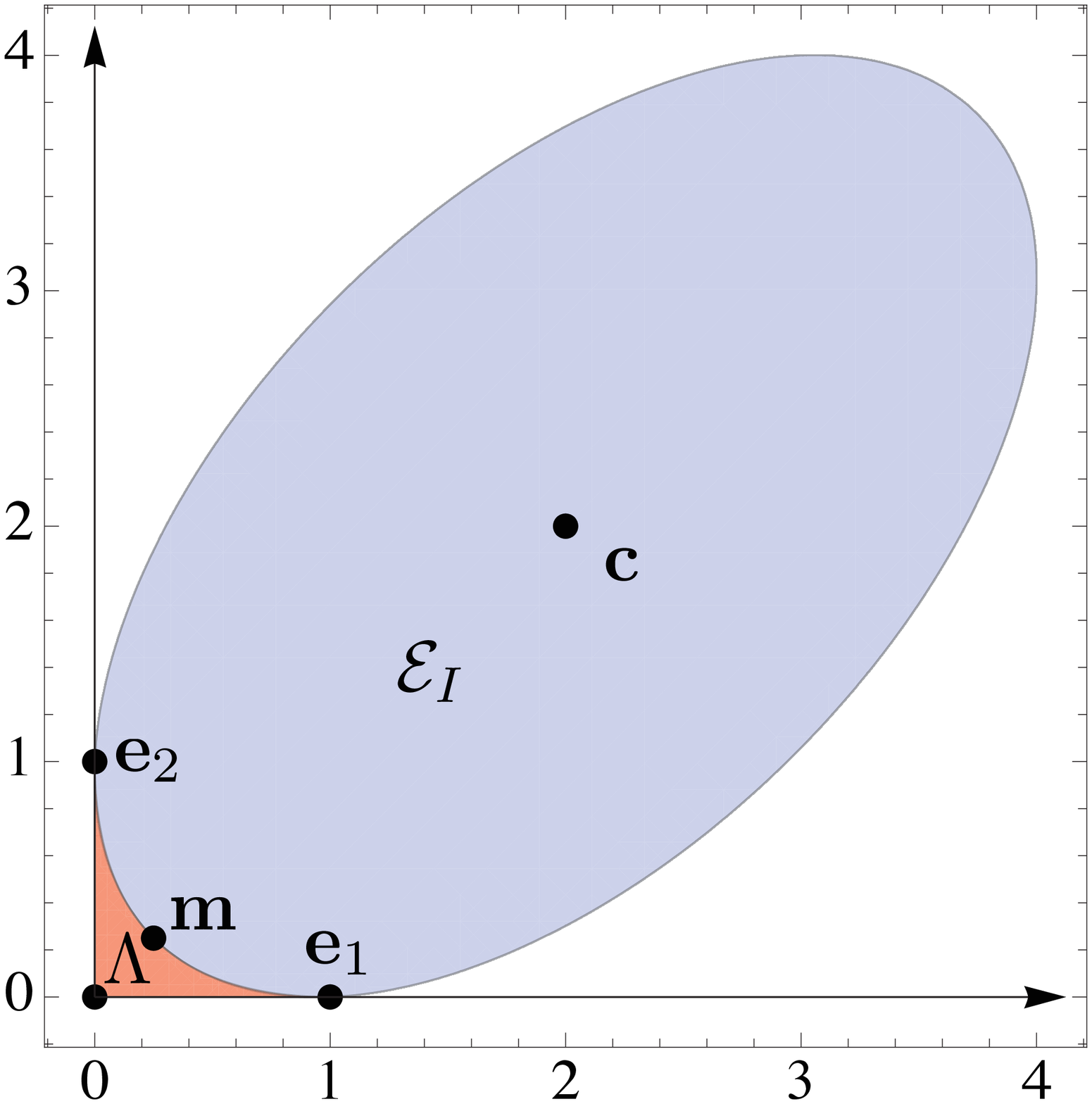}
\includegraphics[width=1.7in]{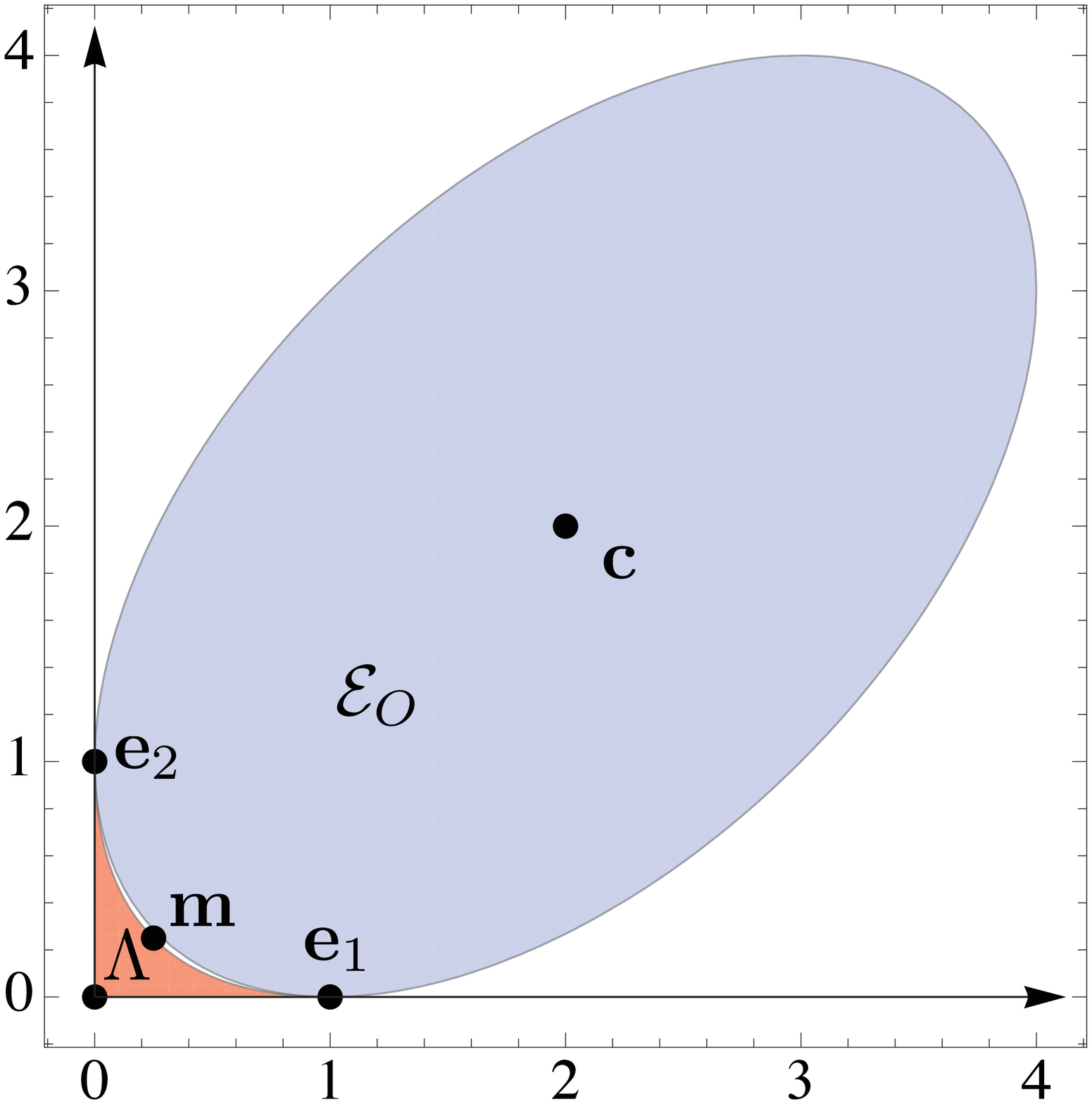}
\caption{The (conjectured) inner (left) and outer (right) bounds on $\Lambda$ using $c=0.55$ (top) and $c=2$ (bottom).  The quality of the bound improves as $c$ increases.}
\label{fig:io}
\end{figure}

Fig. \ref{fig:io} illustrates the ellipsoids $\Emc_I,\Emc_O$ and the associated inner and outer bounds $\Lambda_I,\Lambda_O$ for the special case of $n=2$ for $c=0.55$ and $c=2$.  It is observed that the accuracy of the bounds improves as $c$ is increased.

If the conjectured bounds hold then $\partial \Lambda$ is increasingly tightly ``sandwiched'' between part of $\partial \Emc_{I}$ and $\partial \Emc_{O}$ as $c \rightarrow \infty$.  This sandwiching is asymptotically tight at the all-rates-equal point $\mbf$ (and by construction always tight at each $\ebf_i$).  Although this tightness decreases in $n$ it is bounded.  Specifically, the ellipsoids $\Emc_I,\Emc_O$ viewed in the limit as $c \to \infty$ have axes ratios given by $\lim_{c \to \infty} a_{1,I}^2/a_{1,O}^2 = 1$ and $\lim_{c \to \infty} a_{2,I}^2/a_{2,O}^2 = \frac{1}{2\left(1-\left(1-\frac{1}{n}\right)^{n-1}\right)}$ for each $n$ (see \eqref{eq:EIEO}).  This latter expression equals $1$ when $n=2$ and monotonically decreases in $n$ to $\erm/(2(\erm-1)) \approx 0.791$. Thus when $n=2$, the $\Emc_I,\Emc_O$ are asymptotically equal as $c \to \infty$. For large $c$ and for arbitrary $n$, the axis ratio for $a_1$ is always approximately one, while the axis ratio for $a_2^2$ is bounded between $(0.791,1.0)$.

Fig.~\ref{fig:vol} shows the volumes of $\Lambda, \Lsrs, \Lbc, \Lambda_I, \Lambda_O$ versus $n = 2,\ldots,6$, with $c=100$. Monte-Carlo simulation was used to obtain the volumes of $\Lbc,\Lambda_I,\Lambda_O$, while \eqref{eq:2} and \eqref{eq:u} were used for the volumes of $\Lambda$ and $\Lsrs$, respectively.


\section{Stabilizing controls for a rate vector} 
\label{sec:5} 

\begin{figure}[!htbp]
\centering
\includegraphics[width=2.25in]{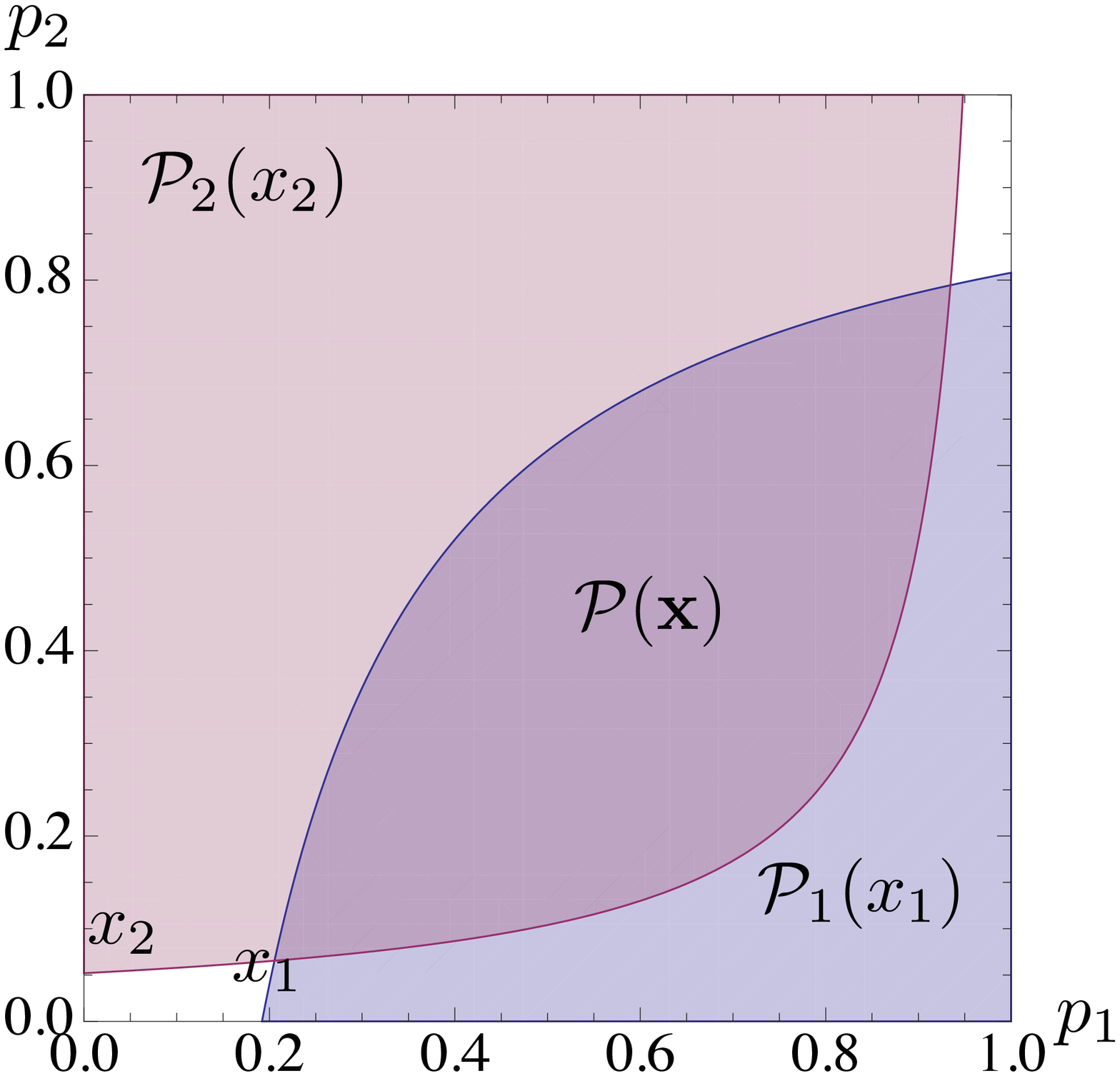}
\caption{The region $\Pmc(\xbf)$ for $\xbf = (0.192,0.052)$.}
\label{fig:pl}
\end{figure}

We now change gears and discuss the set $\Pmc(\xbf)$ in \eqref{eq:t}, which is the set of contention probabilities that can stabilize the given arrival rate vector $\xbf$ assuming worst-case service rates.  The motivation is that knowing a rate vector is stabilizable (i.e., whether or not $\xbf \in \Lambda$ or $\xbf \in \Lambda_A$) is useful only if you also know {\em how} it may be stabilized, i.e., $\Pmc(\xbf)$.  Note $\Pmc(\xbf) \subseteq \Pmc_A(\xbf)$ for $\Pmc_A(\xbf)$ the set of controls that stabilize Aloha under arrival rates $\xbf$.  Our next result establishes that $\Pmc(\xbf)$ is convex.  As an example, the region $\Pmc(\xbf)$ for $\xbf = (0.192,0.052)$ is shown in Fig.~\ref{fig:pl}.  

\begin{proposition}
\label{pro:6}
The set $\Pmc(\xbf)$ in \eqref{eq:t} is convex for all $\xbf \in \Lambda$.  
\end{proposition}

\begin{proofsketch}
We can decompose $\Pmc(\xbf)$ as the intersection of the $n$ regions $\Pmc(\xbf) = \bigcap_{i=1}^n \Pmc_i(x_i)$ where 
\begin{equation}
\Pmc_i(x_i) = \left\{ \pbf : x_i \leq p_i \prod_{j \neq i} (1-p_j) \right\}, ~ i \in [n].
\end{equation}
Since convexity is preserved under intersection, it suffices to show that each $\Pmc_i(x_i)$ is convex.  Letting $\pbf_{-i}$ be the $(n-1)$-vector $\pbf \setminus p_i$ and $f_i(\pbf_{-i}) = x_i / \prod_{j \neq i} (1-p_j)$, we can express
\begin{equation}
\label{eq:s}
\Pmc_i(x_i) = \left\{ (\pbf_{-i},p_i) : f_i(\pbf_{-i}) \leq p_i \right\}, ~ i \in [n].
\end{equation} 
But \eqref{eq:s} is the epigraph for $f_i$ and thus $\Pmc_i(x_i)$ is a convex set iff $f_i$ is a convex function (\cite{BoyVan2004} p.75).  The function 
\begin{equation}
\tilde{f}_i(\pbf_{-i}) \equiv \log f_i(\pbf_{-i}) = \log x_i + \sum_{j \neq i} -\log(1-p_j)
\end{equation}
is a convex function of $\pbf_{-i}$ since $-\log(1-p_j) = - \log(1-\ebf_j^{\Tsf} \pbf_{-i})$ is a convex function of $\pbf_{-i}$ and $\tilde{f}_i$ is therefore a non-negative weighted sum of convex functions, and is therefore itself convex (\cite{BoyVan2004} p.79).  Since $f_i$ is log-convex it is also convex (\cite{BoyVan2004} p.104).  $\blacksquare$
\end{proofsketch}

\vspace{-0.25in}


\section{Conclusion} 
\label{sec:6} 

We have provided easy-to-check inner and outer bounds on the set $\Lambda$, which is a paradigmatic example of an inner bound on the Aloha stability region that is difficult to check membership. There are many directions to pursue to extend this work, most notable is to prove the Conjecture \ref{con:2} (which extends Conjecture \ref{con:1}).  Explicit expressions for the volumes of these sets as a function of $n$ are desirable, but appear difficult.

\vspace{-0.125in}

\bibliographystyle{IEEEtran}
\bibliography{ISIT2010-AlohaBib}
\end{document}